\documentclass[figures]{epl}
\usepackage{amsmath,amssymb,graphicx}

\title{Towards a phase diagram of the 2D Skyrme model}
\author{Oliver Schwindt\thanks{e-mail: \email{schwindt@theory.phy.umist.ac.uk}}
 and Niels R. Walet\thanks{e-mail: \email{Niels.Walet@umist.ac.uk}}}
\institute{Department of Physics, UMIST, PO Box 88, Manchester, M60 1QD,  UK}

\begin{document}

\maketitle

\begin{abstract}
We discuss calculations of the phase diagram of the baby-Skyrme model,
a two-dimensional version of the model that has been so successful in
the description of baryons. Contact is made with the sine Gordon model
in 1D, and relations with the Skyrme model used in the quantum-Hall
effect are pointed out. It is shown that at finite temperature the
phase diagram is dominated by a liquid, and not the crystal that
plays a role for zero temperature.
\end{abstract}
\section{Introduction}

The Skyrme model has a venerable history \cite{Skyrme} in the
non-perturbative description of nucleon structure and the low-energy
behaviour of baryonic matter, since it contains a good description of
the long-wave length behaviour of the dynamics of hadrons. As has been
argued by 't Hooft and Witten \cite{tHooftWitten}, this is closely
related to the large-number-of-colours limit of QCD, in which baryons
must emerge as solitons, in much the same way as happens in the Skyrme
model. Alternatively, we can interpret the model in terms of chiral
perturbation theory, which corresponds closely to a gradient expansion
in terms of the pion field.  The model can be used to describe, with
due care \cite{SkyrmeNuclei}, systems of a few nucleons, and has also
been applied to nuclear and quark matter.  One of the most complicated
aspects of the physics of hadrons is the behaviour of the phase
diagram of hadronic matter at finite density, and low or even zero
temperature. The complex interplay between confinement and chiral
symmetry is difficult to describe, and every model has its own
draws-backs.  Within the standard zero-temperature Skyrme model
description there are signatures of chiral symmetry restoration at
finite density, but in a rather strange way, where a crystal of
nucleons turns into a crystal of half nucleons at finite density,
which is chirally symmetric only on average \cite{Jackson}. 
The question of the finite-temperature behaviour 
has never been addressed and would
be of some interest. Since the behaviour of nuclear matter at high
densities is expected to reveal, probably by one or two phase
transitions, the substructure of baryonic matter, it is interesting to
study this phenomenon in Skyrme models. Before doing so, we would like to study a slightly
simpler form of the model, by reducing the number of dimensions to two.

The two-dimensional Skyrme model also has physical relevance; a
special form of the model is used for ferromagnetic quantum Hall (QHF)
systems \cite{Girvin}. This effective theory is obtained when the
excitations relative to the $\nu=1$ ferromagnetic quantum Hall state
are described in terms of (a gradient expansion in) the spin density,
a field with properties analogous to the pion field in nuclear physics
\cite{LK90}. Apart from obvious changes due to the number of
dimensions, the new approach differs from the historical Skyrme model
by having a different time-dependent term in the Lagrangian, and the
appearance of a non-local interaction, where the topological charge
density at different points interacts through the Coulomb force. In
the limit of large Skyrmions this last term can be approximated by a
more traditional local ``Skyrme term'', which is quartic in the
fields, leading to the standard baby-Skyrme model with local
interactions.  Even with the non-local complications the model is on
the whole remarkably similar to the nuclear Skyrme model. The
Skyrmionic effective degrees of freedom describe the ground state of
such systems, and probably also the low energy dynamics and
thermodynamics, so that we can ask similar questions as for baryonic matter.

It has been argued \cite{BFCD95} that for filling factors not too far
from $\nu=1$ the quantum ground state is described by a crystalline
solution of the Skyrme model, even though no direct experimental proof
exists. The most exciting possible experimental signature is a low-temperature
phase transition, which may be associated with Skyrmion-lattice
melting
\cite{TGF98}. This last paper also makes an important study of the
possible phases of the quantum Hall Skyrme model. Like in most other
work on the subject
\cite{AE98,MM00} these calculations are based on an assumption that
the Skyrmions in a crystal are identical to free Skyrmions, which is
at best a crude approximation of the real situation. 
%It has also been
%argued \cite{BP99} that quantum fluctuations may actually destroy
%the crystaline structure, perhaps the only question that can not be addressed
%by the approach in this letter.
At the same time it allows us to look at the phase diagram of quantum
Hall Skyrmionic systems, which may well be the easiest way to
calculate parts of the phase diagram of the underlying electronic
model.

In this paper we shall consider the local baby Skyrme model, which can
be seen as a dimensional reduction of the 3D Skyrme model, or as the
local limit of the QHF Skyrme model.  It is also related to the
sine-Gordon model in 1D, by a simple dimensional reduction. Thus all
numerical techniques can first be tested on this simple and exactly
solvable model.  All of these models contain a topologically conserved
charge, which gives rise to topological solitons. In 3D (nuclear
physics) these are identified as baryons, and in QHF systems as
quasi-particles.

At finite density but zero temperature the Skyrme models
and the sine Gordon model have a crystalline structure, which consist
of regularly spaced solitons. In the one-dimensional case, where exact
solutions exist \cite{SG}, this is known to be a zero temperature
artifact, and we have a liquid at any finite temperature, no matter
how small. In the nuclear Skyrme model we would like to find a fluid,
to mimic the quantum liquid behaviour expected in nuclear matter
\cite{Cohen}.  In order to appreciate the subtleties involved, one
must understand that the Skyrme model, as a classical field theory,
can be understood as a semiclassical (large action) limit of a quantum
theory.  Clearly the quantum fluctuations in the underlying theory
could be large enough to wash out the crystalline structure, as happens
for the sine Gordon model. In the special case of one space dimension
it is also well known that thermal and quantum fluctuations play
exactly the same role, and thermal fluctuations also break the
crystalline state. For the Skyrme and baby Skyrme models it is not easy
to access the quantum fluctuations, since the field theories are
non-renormalisable, but we can access the thermal fluctuations (with
care, because nonremormalisability plays a role in the thermodynamics
as well!).  Furthermore, we might well wish to study the physics of
these systems at temperatures where thermal fluctuations dominate the
physics. Therefore, we shall concentrate on the finite temperature
phases of these classical field theories.

\section{Theoretical and computational background}

The most direct approach to the problem is to perform a Monte-Carlo
(Metropolis algorithm based) study of the partition function of each
of these models.  As we shall argue below the least obvious aspect in
such an approach is how to deal with the topological conservation
laws. Rather than immediately tackle the 3D model, we concentrate on
the (local) 2D model as a first example, for which the visualization
of, and thus the understanding gained from, the results is much more
straightforward.  The extensions to 3D (nuclear physics) and
quantum-Hall Skyrmions are under way and will be presented elsewhere
\cite{Oli2,Weidig}. We expect the current for the QHE solitons to be very 
similar to the results reported here.

The model is defined by the Lagrange density, which in $D$ dimensions
takes the form (we shall work in ``natural'' units in which
the parameters $\alpha$ and $\beta$ equal 1)
\begin{eqnarray}
{\cal L} &=& \frac{\alpha}{2} \sum_{k,\mu}(\partial_\mu \phi_k)^2
-\frac{\beta}{4} \left(\sum_{k,\mu}(\partial_\mu \phi_k)^2\right)^2
%\nonumber\\&&
-\frac{1}{16} \sum_{k,l}\Bigl(\sum_\mu\partial_\mu \phi_k \partial_\mu \phi_l\Bigr)^2
%\nonumber\\&&
-m^2_\pi (1-\phi_1),\label{eq:L}
\end{eqnarray}
where the coordinates $\mu$ run from $0$ (time) to $D$, and the
vector field $\phi$  has $D+1$ components, and has unit length
\begin{equation}
\sum_k\phi_k^2=1.
\end{equation}

This form of Lagrangian is the exact analogue of the original nuclear
(3D) one, apart from the ``mass term'' which was first added to the 3D
model by Adkins {\em et al} \cite{ANW} to improve the physics. In the
2D model it is crucial for the stability of the solitonic solutions;
in the QHE eefct such a term natural arrises from the Zeeman effect.
Nonetheless the current model differs in several important aspects
from the quantum-Hall Skyrme model, the main differences are the
fourth-order term, which is non-local in the QHE, and the second order
time derivatives, where only a first order-one appears in the
Lagrangian of the QHE (see, however, Ref.~\cite{Stoof}). Nonetheless,
it might be expected that the current model shares many of the
characteristics of the 3D model and the QHE Skyrmions, and at least
the zero-temperature results are independent of the term containing
time-derivatives.

Apart from the question what the few-soliton solutions of the model
look like \cite{HSW}, one might ask what phases are shown by the
model at finite density and temperature. 
The way we shall address this problem is by using a Metropolis
approach, where the partition function is sampled using a pseudo-time
dynamics.  We first discretize space on a lattice, and then calculate
the partition function,
\begin{equation}Z_N(\beta)=
\int\prod_{i}
d^{3}\phi_{i}\delta(\vec{\phi}_i^2-1)
d^{3}\dot{\phi}_{i}\delta(\dot{\vec{\phi}}_{i}\cdot \vec{\phi}_{i})
\exp(-\beta E),\label{eq:Z}
\end{equation}
where $i$ runs over all lattice sites.  Here $E$ is the energy
expression corresponding to the Lagrangian (\ref{eq:L}), and the delta
functions arise from the unitarity constraint and its time derivative.
Since we cannot work with an infinite system, we use a finite box with
periodic boundary conditions. This then raises the final question, why
we have chosen to use a canonical (rather than grand-canonical)
ensemble in the expression above. The reason for this choice combines
practical limitations with topological ones: We have found it
impossible to design a sufficiently simple (i.e., localised) update
mechanism in the Metropolis algorithm that succeeds in generating
additional topological charge. This is clearly related to the
non-local nature of the topological conservation laws, and we find
that the charge is essentially constant over the whole Monte-Carlo
iteration, even if we add an additional chemical potential term $\mu
B$ to the exponential in Eq.~(\ref{eq:Z}). Since the total charge is
thus fixed by the initial one, our approach is rather ineffective in
sampling the grand-canonical ensemble. In order to sample this fully,
we must superimpose the results from various canonical simulations
with different initial baryon number.  Alternatively we have found it
convenient to study systems with open boundaries, an approach not
pursued in this work. Such an approach may show even more promise when
combined with a heat bath attached to those boundaries, or related
Langevin calculations, see e.g.~\cite{arab}.

\section{Results}

\begin{figure}
\twofigures[width=6.5cm]{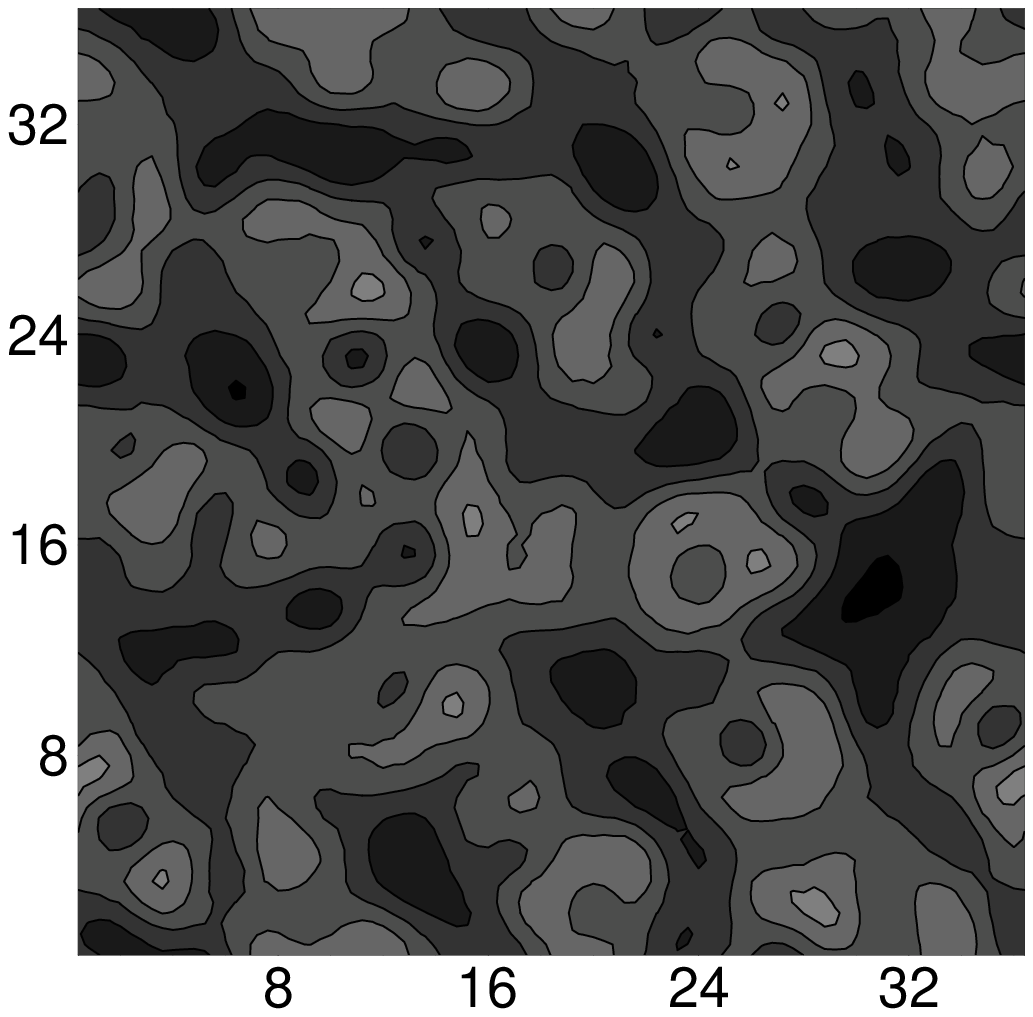}{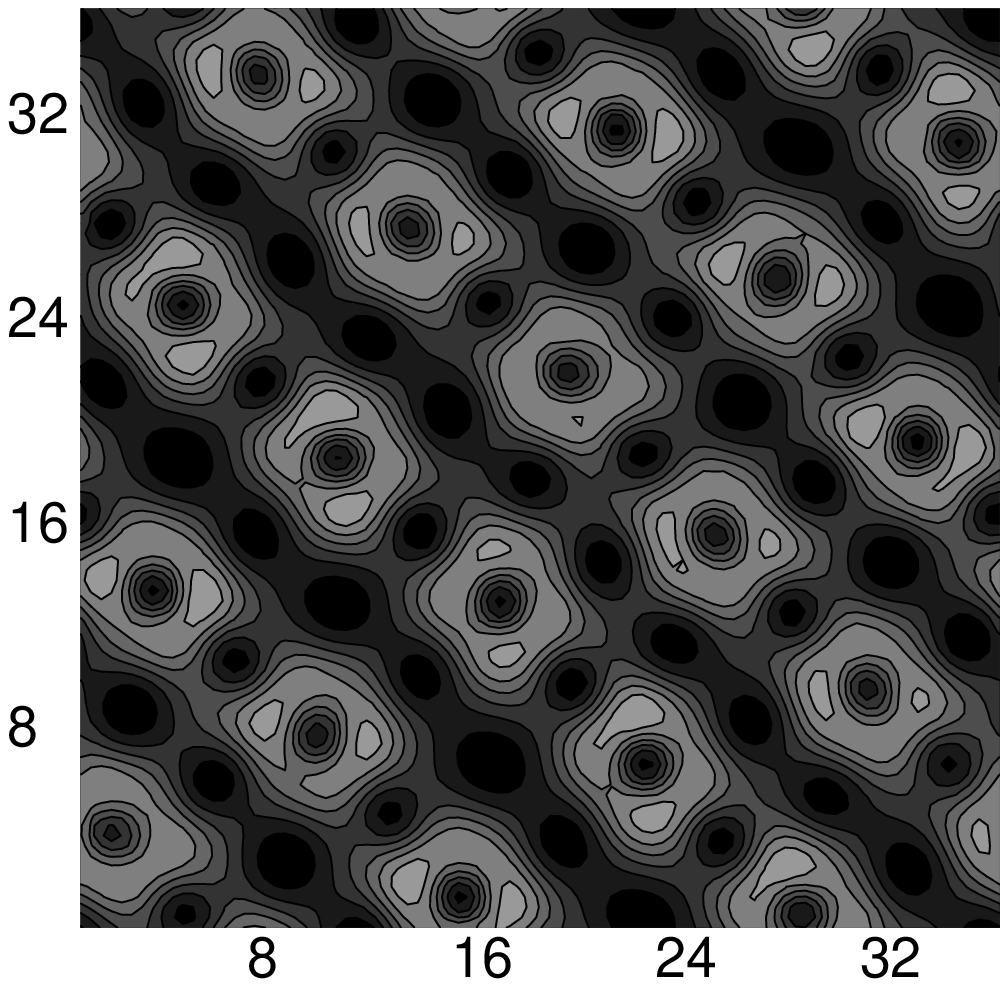}
\caption{The average of the baryon density for a fluid phase.
The number of simulations included in the average was chosen  small, do that some
structure is still visible.}\label{fig:fluid1}
\caption{The ensemble average of the baryon density for a solid.}\label{fig:solid}
\end{figure}

The Monte-Carlo simulations were verified by comparison with exact
results for the sine-Gordon model \cite{SG}, and found to be quite accurate, up
to truncation errors due to the lattice spacing. Full results shall be
presented in Ref.~\cite{Oli2}. 
%We then investigated the crystalline phases of the baby Skyrme model. 

Our main goal is to understand the thermodynamics, and the order in
the system as a function of density $\rho$ and temperature $T$.  To
that end we need to be able to distinguish fluid phases from solid
ones. The most visual way is the pseudo-time averaged topological
density. For a fluid, which its large mobility, such an average is
constant, see Fig.~\ref{fig:fluid1}. In order to show strucuture we
have limited the number of configurations used in the average
so that some of the structure in  the individual snapshots 
(field configurations) contributing to the average is still visible.

This needs to be contrasted to the behaviour of a solid, where
 we find that the snap-shots and the time averages show similar
structures, but the fluctuations disappear in the averages, and we find
a very clear solid structure, see Fig.~\ref{fig:solid}.
\begin{figure}
\begin{center}
\includegraphics[width=7cm]{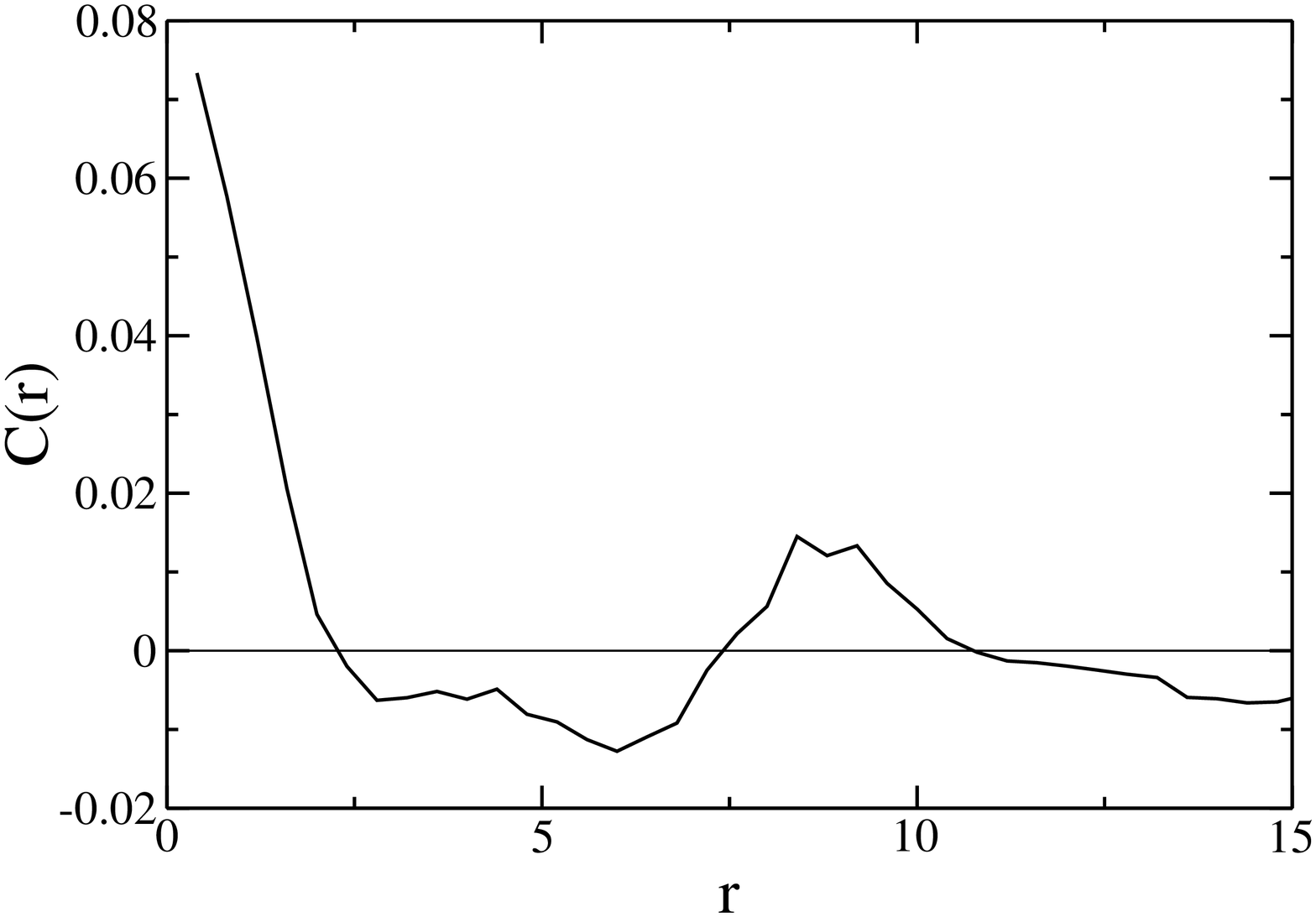}\includegraphics[width=7cm]{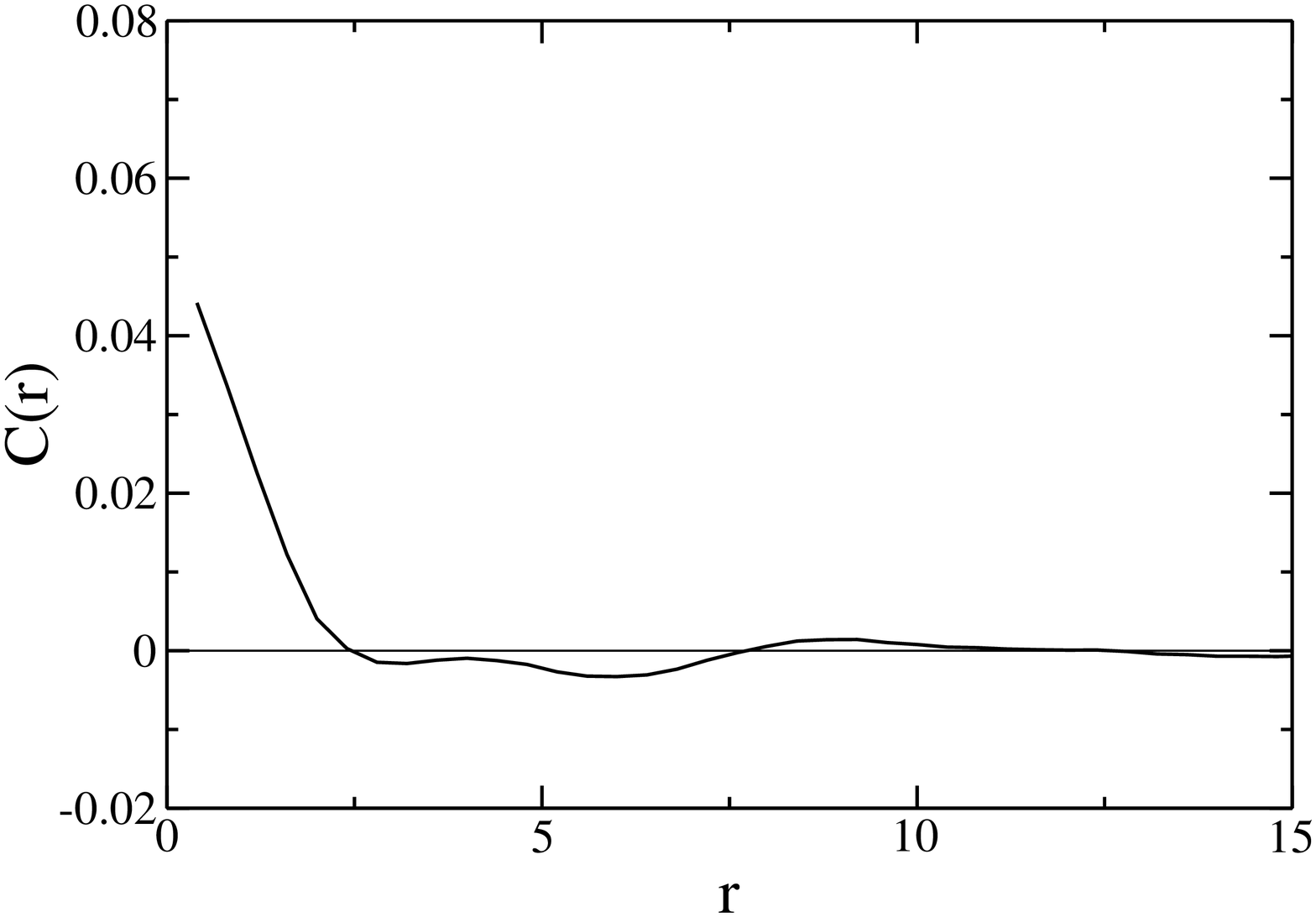}
\end{center}
\caption{Correlation functions for a solid (left), and a liquid (right).}
\label{fig:correlate}
\end{figure}
The identification of the crystalline structure can also be made by
looking at the long-range order exhibited by the baryon-density correlation
function
\begin{equation}
C(\vec r)=\bigl\langle \int d^2r' B(\vec r') B(\vec r'-\vec r) \bigr\rangle/B^2,
\end{equation}
where $B$ is the total baryon number. We have chosen to calculate the
angle average $C(r)$, which should give all details of the
correlations for a fluid, and contains partial information in the
case of a solid.  The behaviour for a solid is quite different from
that of a liquid, as we can see in Fig.~\ref{fig:correlate}. 
The liquid shows some correlations, which is of course the
distinction between liquid and gas, and shows that our identification
is probably correct, even though we cannot be absolutely certain.

\begin{figure}
\begin{center}
\includegraphics[width=7cm]{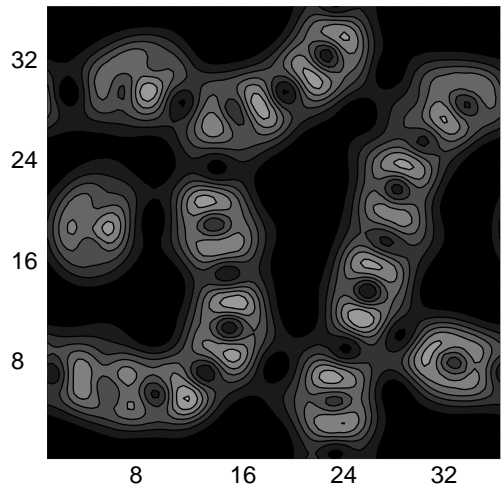}
\end{center}
\caption{Ensemble averaged baryon density in the phase coexistence region.}
\label{fig:coexistance}
\end{figure}

Unfortunately that is not the full story. We find no indication of a
phase transition to a gas phase, but we do find something that can
best be called phase coexistence, see Fig.~\ref{fig:coexistance}. We
find long solid-like networks (or, in other words, complicated
multisolitons) and empty areas. We cannot be completely sure about
their meaning, but the fact that this occurs for low densities may indicate
some kind of droplet or grain formation.

\begin{figure}
\begin{center}
\includegraphics[width=7cm]{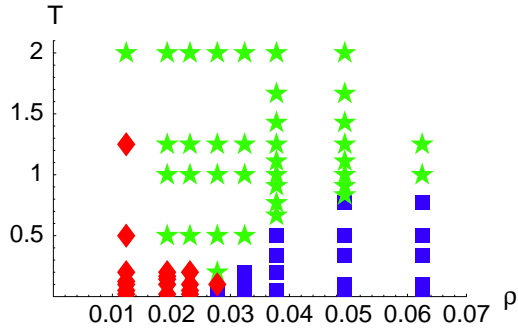}
\end{center}
\caption{The phase diagram for the baby Skyrme model.
Squares denote a solid, stars a liquid and lozenges phase coexistence.
}\label{fig:phasediag}
\end{figure}

We have performed numerous simulations, in order to get a more
detailed picture of what is going on. The resulting phase diagram is
shown in Fig.~\ref{fig:phasediag}. We find that at finite temperatures
a large part of the phase diagram is indeed dominated by a fluid
phase, which gives us cause to investigate the traditional Skyrme
model in more detail.

\section{Conclusions and Outlook}
\begin{figure}
\begin{center}
\includegraphics[width=7cm]{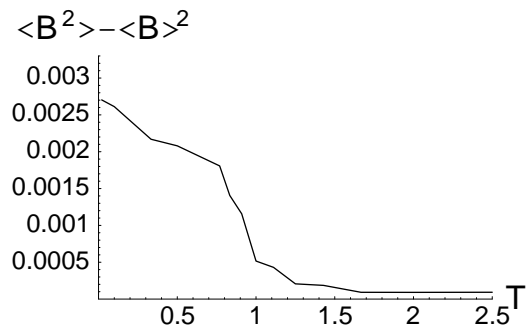}
\end{center}
\caption{The fluctuations in baryon density for fixed $\rho$.
}\label{fig:Bfluc}
\end{figure}
We have not yet addressed the question what the nature of the nature
of the phase transition is. Our results are not yet of sufficient
quality to make a definite pronouncement on this issue.  There are
various indications that the phase transition is of low, maybe even
first order. We have considered the energy as a function of $\rho$ and
$T$. We find indications of a sudden jump in flucutuations of the
baryon density, see Fig.~\ref{fig:Bfluc}, at the phase transition, and
a rapid change in energy. We are unable to state whether the change in
the energy has a discontinuous slope; that will have to be
investigated seperately. We also found an indication of latency (i.e.,
the point where the phase transition occurs depends on the direction
in which we approach it). This usually indicates a first order phase
transition, but the question is worthy of further investigation.

Of more immediate interest is the application to more realistic
models, such as the 3D Skyrme model, or the QHE Skyrmions. Both of
these are much more computationally intense problems, but we are
currently making some inroads into both of these areas, using
techniques similar to the ones reported here. We expect to be able to
present some results in the near future \cite{Oli2,Weidig}.

\acknowledgments This work was supported by research grants
(GR/L22331 and GR/N15672) from the Engineering and Physical Sciences
Research Council (EPSRC) of Great Britain.  OS has also been supported
by an EPSRC studentship.

\end{document}